\documentclass[aps,showpacs,showkeys,pre,twocolumn,superscriptaddress,floatfix]{revtex4}
 \usepackage{graphicx,bm,amssymb,amsmath,dcolumn,hyperref}
 \usepackage{multirow}
 \usepackage{color}
 \usepackage{subfigure}  
 \usepackage{epstopdf}
 \usepackage{bbm}
 \usepackage{graphicx}
 \usepackage{hyperref}
 \usepackage{threeparttable}
 \usepackage{amsthm}

\begin{document}
\newcommand{\be}{\begin{equation}}
\newcommand{\ee}{\end{equation}}
\newcommand{\half}{\frac{1}{2}}
\newcommand{\ith}{^{(i)}}
\newcommand{\im}{^{(i-1)}}
\newcommand{\gae}
{\,\hbox{\lower0.5ex\hbox{$\sim$}\llap{\raise0.5ex\hbox{$>$}}}\,}
\newcommand{\lae}
{\,\hbox{\lower0.5ex\hbox{$\sim$}\llap{\raise0.5ex\hbox{$<$}}}\,}

\definecolor{blue}{rgb}{0,0,1}
\definecolor{red}{rgb}{1,0,0}
\definecolor{green}{rgb}{0,1,0}
\newcommand{\blue}[1]{\textcolor{blue}{#1}}
\newcommand{\red}[1]{\textcolor{red}{#1}}
\newcommand{\green}[1]{\textcolor{green}{#1}}

\newcommand{\scrA}{{\mathcal A}}
\newcommand{\scrL}{{\mathcal L}}
\newcommand{\scrN}{{\mathcal N}}
\newcommand{\scrS}{{\mathcal S}}
\newcommand{\scrs}{{\mathcal s}}
\newcommand{\scrP}{{\mathcal P}}
\newcommand{\scrO}{{\mathcal O}}
\newcommand{\scrR}{{\mathcal R}}
\newcommand{\scrC}{{\mathcal C}}

\newcommand{\dm}{d_{\rm min}}
\title{Simultaneous analysis of three-dimensional percolation models}
\date{\today}
\author{Xiao Xu}
\affiliation{Hefei National Laboratory for Physical Sciences at Microscale and Department of Modern Physics, University of Science and Technology of China, Hefei, Anhui 230026, China}
\author{Junfeng Wang}
\affiliation{Hefei National Laboratory for Physical Sciences at Microscale and Department of Modern Physics, University of Science and Technology of China, Hefei, Anhui 230026, China}
\affiliation{School of Electronic Science and Applied Physics, Hefei University of Technology, Hefei, Anhui 230009, China}
\author{Jian-Ping Lv}
\email{phys.lv@gmail.com}
\affiliation{Department of Physics, China University of Mining and Technology, Xuzhou 221116, China}
\author{Youjin Deng}
\email{yjdeng@ustc.edu.cn}
\affiliation{Hefei National Laboratory for Physical Sciences at Microscale and Department of Modern Physics, University of Science and Technology of China, Hefei, Anhui 230026, China}

\begin{abstract}
We simulate the bond and site percolation models on several three-dimensional lattices,
including the diamond, body-centered cubic, and face-centered cubic lattices.
As on the simple-cubic lattice [Phys. Rev. E, \textbf{87} 052107 (2013)],
it is observed that in comparison with dimensionless ratios based on cluster-size distribution,
certain wrapping probabilities exhibit weaker finite-size corrections and are more sensitive
to the deviation from percolation threshold $p_c$, and thus provide a powerful means for determining $p_c$.
We analyze the numerical data of the wrapping probabilities
simultaneously such that universal parameters are shared by the aforementioned models, and
thus significantly improved estimates of $p_c$ are obtained.
\end{abstract}
\maketitle

\section{Introduction}
Percolation is a geometric model which involves the random occupation of sites or edges of a regular lattice, and was first
introduced by Broadbent and Harmmersley~\cite{BroadbentHarmmersley57}. As a cornerstone of the theory of critical
phenomena~\cite{StaufferAharony94} and a central topic in probability theory~\cite{Grimmett99,BollobasRiordan09},
percolation attracts much attention.

The two-dimensional (2D) case has been studied extensively,
and several exact results are known. Coulomb gas
arguments~\cite{Nienhuis87} and conformal field theory~\cite{Cardy87}
predict the exact values of the bulk critical exponents
$\beta =5/36$ and $\nu=4/3$, which have been confirmed rigorously
in the specific case of site percolation on the triangular
lattice~\cite{SmirnovWerner01}.
Moreover, percolation thresholds $p_c$ on many 2D lattices are
exactly known~\cite{Essam72}, or known to very high precision~\cite{Feng08,DingPRE10}.
For $d>2$, estimates of $p_c$ have to rely on numerical methods such as series expansions
and Monte Carlo simulations, while the critical exponents $\beta=1$ and $d \nu=3$
for $d\geq d_c =6$ can be predicted by mean-field theory~\cite{Toulouse74} and even proved
rigorously~\cite{AizenmanNewman84,HaraSlade90} for $d\ge19$.
A more or less thorough list of percolation thresholds for $d \in [2,13]$
is summarized on the Wikipedia webpage: http://en.wikipedia.org/wiki/Percolation\_threshold.

Very recently, two of the authors and coworkers carried out an extensive simulation of bond and site percolation
on the simple-cubic (SC) lattice up to system size $512 \times 512 \times 512$~\cite{WangZhouZhangTimDeng13},
and determined the percolation thresholds and critical exponents to high precision.
It was observed that in comparison with dimensionless ratios
based on cluster-size moments, the wrapping probabilities suffer from weaker finite-size corrections and
are more sensitive to the deviation $p-p_c$ from the percolation threshold.
As an extension of Ref.~\cite{WangZhouZhangTimDeng13}, the present work studies percolation on
other common three-dimensional (3D) lattices, and shows that
such an observation generally holds in 3D percolation.
Meanwhile, with the employment of a simultaneous fitting procedure developed in Ref.~\cite{DengBlote03}
and the help of the accurate data reported in Ref.~\cite{WangZhouZhangTimDeng13},
we also provide high-precision estimates of $p_c $ for the site and bond percolation on the diamond (DM),
body-centered cubic (BCC), and face-centered cubic (FCC) lattices.

The remainder of this paper is organized as follows.
Section~\ref{Model and simulation approach}
defines the sampled quantities of interest.
In Sec.~\ref{estimating pc},
the numerical data of the dimensionless ratios and the wrapping probabilities are
analyzed separately for each percolation model. Then,
a simultaneous fitting of the wrapping probabilities is carried out to determine percolation threshold $p_c$.
Section~\ref{results at pc} presents the analyses for other quantities at criticality $p_c$,
and a brief discussion is given in Sec.~\ref{discussion}.

\section{Sampled quantities}
\label{Model and simulation approach}
\begin{figure}
\vspace*{0cm} \hspace*{-0cm}
\begin{center}
\includegraphics[width=0.35\columnwidth]{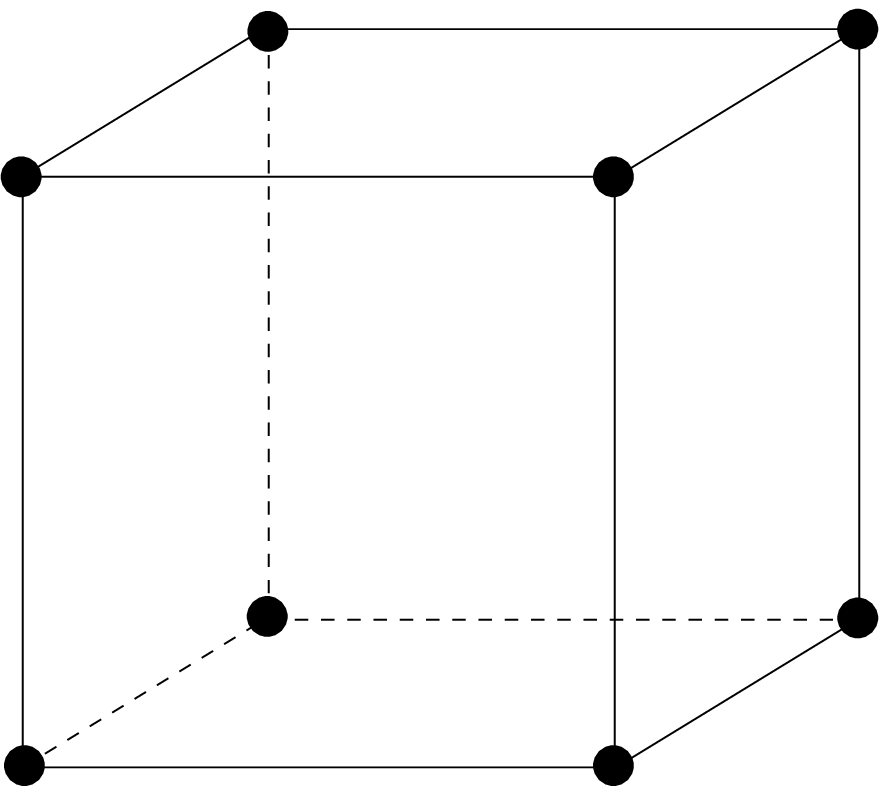}
\includegraphics[width=0.35\columnwidth]{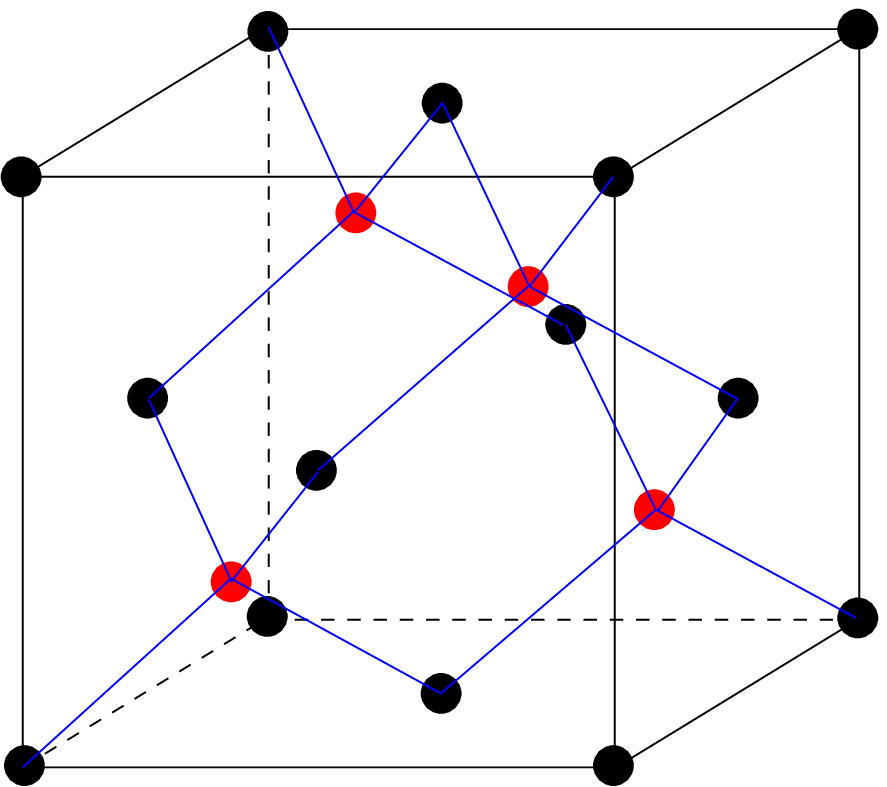} \\
\hspace*{2mm}
\includegraphics[width=0.35\columnwidth]{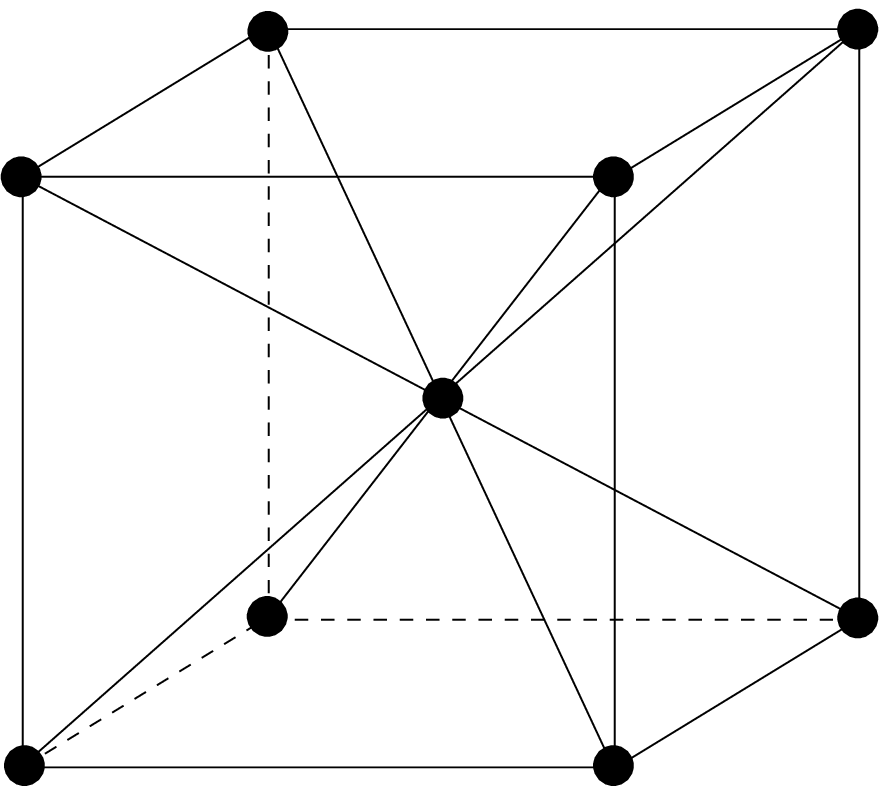}
\includegraphics[width=0.35\columnwidth]{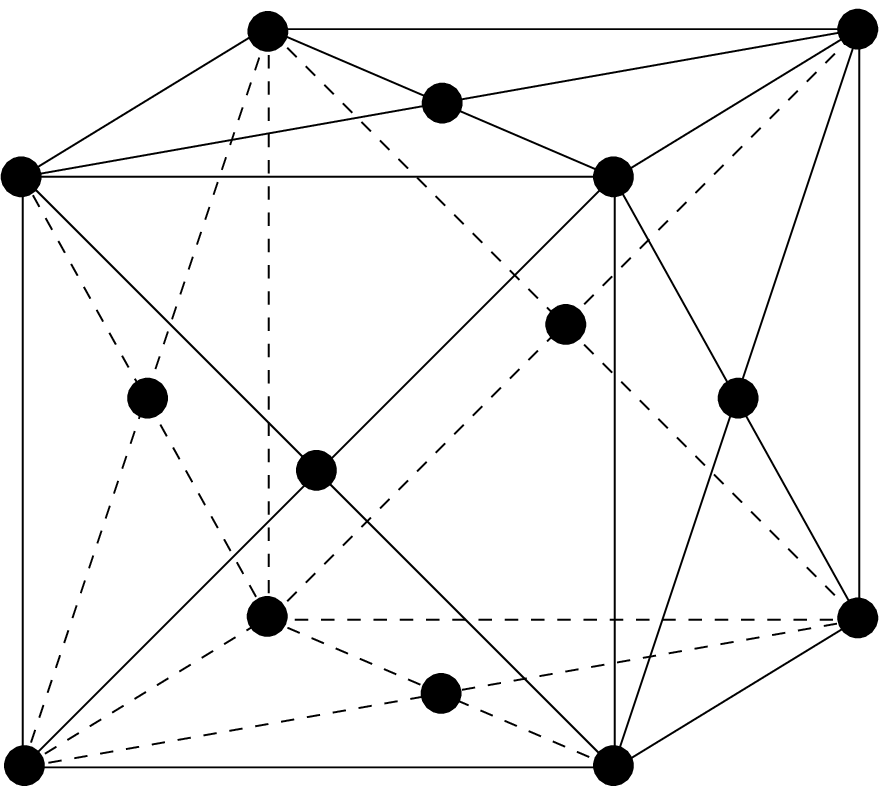} \\
\hspace*{2mm}
\end{center}
\vspace*{-10mm}
\caption{Three-dimensional lattices: (left-top), SC; (righttop),
DM; (left-bottom), BCC; (right-bottom), FCC.}
\label{fig1}
\end{figure}

We study bond and site percolation on three-dimensional lattices including the
DM, SC, BCC, and FCC lattices,
illustrated in Fig.~\ref{fig1}. The simulations follow the standard method:
each edge/site is occupied with probability $p$ and clusters are constructed by
the breadth-first search.

The sampled quantities are the same as in Ref.~\cite{WangZhouZhangTimDeng13}.
For completeness, they are described in the following.
 \begin{itemize}
 \item The number of occupied bonds $\scrN_b$ or sites $\scrN_s$.
 \item The number of clusters $\scrN_c$.
 \item The largest-cluster size $\scrC_1$.
 \item The cluster-size moments $\scrS_m  = \sum_{C}|C|^m$ with $m=2,4$,
  where the sum runs over all clusters $C$ and $|C|$ denotes cluster size.
 \item An observable $\scrS :=\max\limits_{C}\,\max\limits_{y\in C}\,d(x_C,y)$ used to determine the shortest-path exponent.
   Here $d(x,y)$ denotes the graph distance from vertex $x$ to vertex $y$, and $x_C$ is the vertex in cluster $C$
  with the smallest vertex label, according to some fixed (but arbitrary) vertex labeling.
 \item The indicators $\scrR^{(x)}$, $\scrR^{(y)}$, and $\scrR^{(z)}$, for the event that
  a cluster wraps around the lattice in the $x$, $y$, or $z$ directions, respectively.
 \end{itemize}

 From these observables we calculated the following quantities:
 \begin{itemize}
 \item The mean size of the largest cluster $C_1 = \langle \scrC_1 \rangle$,
       which scales as $C_1\sim L^{y_h}$ at $p_c$, with $L$
       the linear system size and  $y_h = d - \beta/\nu$.

 \item The cluster density $\rho = \langle \scrN_c \rangle/V$, where $V = g L^3$ is the
       number of lattice sites, with $g=1$ for the SC and DM lattices, $g=2$ for the BCC
       lattice, and $g=4$ for the FCC lattice.
 \item The dimensionless ratios
   \begin{equation}
     Q_1  = \frac{\langle {\scrC_1}^2\rangle}{\langle \scrC_1\rangle^2}\;,\;\;\;
     Q_2  = \frac{\langle {\scrS_2}^2\rangle}{\langle 3{\scrS_2}^2 - 2\scrS_4\rangle}\;.
     \label{eq:R}
   \end{equation}
    In the case of the Ising model, $Q_2$ is identical to the dimensionless 
    ratio $Q_M=\langle M^2 \rangle ^2 / \langle M^4 \rangle$,
    where $M$ represents the magnetization.
 \item The mean shortest-path length $S=\langle \scrS \rangle$,
 which at $p_c$ scales like $S \sim L^{\dm}$ with $\dm$ the shortest-path fractal dimension.
 \item The wrapping probabilities
 \begin{equation}
   \begin{aligned}
     R^{(x)} = & \langle \scrR^{(x)} \rangle = \langle \scrR^{(y)} \rangle = \langle \scrR^{(z)} \rangle \;, \\
     R^{(a)} = & 1 - \langle (1-\scrR^{(x)})(1-\scrR^{(y)})(1-\scrR^{(z)})\rangle \;, \\
     R^{(3)} = & \langle \scrR^{(x)}\scrR^{(y)}\scrR^{(z)}\rangle\;.
   \end{aligned}
 \end{equation}
 Here $R^{(x)}$, $R^{(a)}$ and $R^{(3)}$ give the probability that a winding exists in the $x$ direction,
 in at least one of the three possible directions, and simultaneously in the three directions, respectively.
 At $p_c$, these wrapping probabilities take non-zero universal values in the thermodynamic limit $L \rightarrow \infty$.
 \item The covariance of $\scrR^{(x)}$ and $\scrN_b$
 \begin{equation}
   g^{(x)}_{bR} = \langle \scrR^{(x)} \scrN_b\rangle - \langle \scrR^{(x)}\rangle \langle \scrN_b \rangle \; ,
 \label{eq:g}
 \end{equation}
 which scales as $g^{(x)}_{bR}\sim L^{y_t} = L^{1/\nu}$ at criticality $p_c$.
 Analogously, one defines $g^{(x)}_{sR}$ for site percolation, with $\scrN_b$ being replaced with $\scrN_s$.
 \end{itemize}

\begin{figure}
\centering
\includegraphics[scale=0.8]{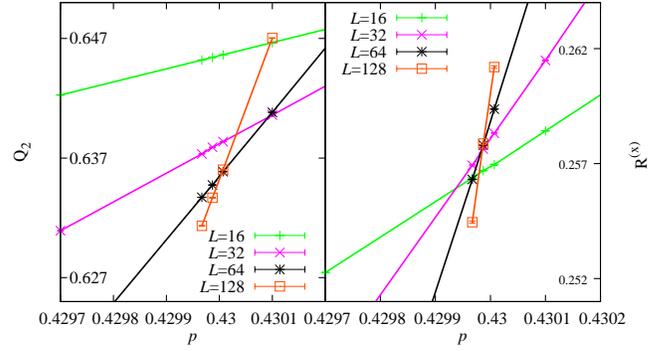}
\caption{Quantities $Q_2$ and $R^{(x)}$ as a function of $p$ for the
site percolation on the DM lattice with various sizes. 
In comparison with $Q_2$, the plot of $R^{(x)}$ has a finer vertical scale, 
but still displays a clearer intersection. 
This suggests that $R^{(x)}$ suffers weaker finite-size corrections and 
provides a better estimator for $p_c$.} 
\label{RxQs}
\end{figure}

\section{Percolation threshold} \label{estimating pc}
The simulation on the SC lattice is up to linear size $L_{\rm max}=512$,
and the number of samples is about $5 \times 10^8$ for $L \leq 128$, $6 \times 10^7$ for $L=256$,
and $3 \times 10^7$ for $L=512$.
The Monte Carlo data and the analysis have been reported in Ref.~\cite{WangZhouZhangTimDeng13}.
For the other lattices, the simulation is less extensive with $L_{\rm max}=128$.
The number of samples is about $10^8$ for lattice $L<128$ and $4\times 10^7$ for $L=128$.

 \begin{table*}
 \begin{center}
 \caption{Percolation thresholds from the separate fits of the wrapping probabilities and the dimensionless ratios.}
 \label{Tab:sep-fit1}
 \scalebox{1.0}{
 \begin{tabular}[t]{|l|l|l|l|l|l|}
 \hline
           & $Q_1$    & $Q_2$   & $R^{(x)}$    & $R^{(a)}$    & $R^{(3)}$   \\
 \hline
  {$\rm{DM}^b$} & 0.389\,591(2) & 0.389\,592(1)  & 0.389\,589\,2(5)  & 0.389\,588\,9(4)  & 0.389\,590\,0(5)  \\
 \hline
  {$\rm{DM}^s$} & 0.429\,987(2) & 0.429\,985(1) & 0.429\,987\,7(9)  & 0.429\,987\,5(6)  & 0.429\,987\,3(4)  \\
 \hline
  {$\rm{SC}^b$}  & 0.248\,811\,96(6)  & 0.248\,811\,92(6) & 0.248\,811\,85(3) & 0.248\,811\,80(4)  & 0.248\,811\,81(9) \\
 \hline
  {$\rm{SC}^s$}  & 0.311\,606\,9(2) & 0.311\,607\,1(2) & 0.311\,607\,68(7) & 0.311\,607\,74(6) & 0.311\,607\,7(1) \\
\hline
  {$\rm{BCC}^b$} & 0.180\,287\,8(9) & 0.180\,288\,3(6) & 0.180\,287\,5(2) & 0.180\,287\,4(2) & 0.180\,287\,9(2) \\
\hline
  {$\rm{BCC}^s$} & 0.245\,961\,7(3) & 0.245\,961\,5(2) & 0.245\,961\,7(2) & 0.245\,961\,70(11) & 0.245\,961\,7(3) \\
\hline
  {$\rm{FCC}^b$} & 0.120\,163\,9(5) & 0.120\,163\,3(3) & 0.120\,163\,6(2) & 0.120\,163\,6(2) & 0.120\,163\,7(3) \\
\hline
 {$\rm{FCC}^s$}  & 0.199\,235\,3(3) & 0.199\,235\,2(2) & 0.199\,235\,2(2) & 0.199\,235\,14(11) & 0.199\,235\,0(2) \\
 \hline
  \end{tabular}
}
 \end{center}
 \end{table*}

 \begin{table}
 \begin{center}
 \caption{Value of the amplitudes $q_1$ obtained from the separate fits of the wrapping probabilities and the dimensionless ratios.}
 \label{Tab:sep-fit2}
 \scalebox{1.0}{
 \begin{tabular}[t]{|l|l|l|l|l|l|}
 \hline
           & $Q_1$    & $Q_2$   & $R^{(x)}$    & $R^{(a)}$    & $R^{(3)}$   \\
 \hline
  {$\rm{DM}^b$} & 0.277(2) & 0.642(3)  & 0.906(4)  & 1.236(5)  & 0.484(3)  \\
 \hline
  {$\rm{DM}^s$} & 0.193(2) & 0.458(4) & 0.652(3)   & 0.894(4)  & 0.341(3)  \\
 \hline
  {$\rm{SC}^b$} & 0.30(3)  & 0.90(7)  & 1.20(7)    & 1.80(9)   & 0.65(7) \\
 \hline
  {$\rm{SC}^s$}  & 0.22(2) & 0.52(4) & 0.70(4) & 1.00(3) & 0.36(3) \\
\hline
  {$\rm{BCC}^b$} & 0.644(3) & 1.46(2) & 2.084(8) & 2.82(3) & 1.12(2) \\
\hline
  {$\rm{BCC}^s$} & 0.30(1) & 0.72(3) & 1.04(2) & 1.42(2) & 0.56(1) \\
\hline
  {$\rm{FCC}^b$} & 1.19(9) & 2.77(4) & 3.91(3) & 5.29(2) & 2.08(2) \\
\hline
 {$\rm{FCC}^s$}  & 0.449(3) & 1.044(9) & 1.507(5) & 2.080(6) & 0.794(4) \\
 \hline
  \end{tabular}
}
 \end{center}
 \end{table}

 \begin{table*}
 \begin{center}
 \caption{Percolation thresholds and other non-universal parameters
  from the simultaneous fits of the wrapping probabilities. For all the fits,
  we set $L_{\rm min} = 32$ for $R^{(x)}$ and $R^{(3)}$ and $L_{\rm min} = 24$ for $R^{(a)}$,
  $a$, $b_1$ and $b_2$ are defined in Eq.~(\ref{eq:simfit}).}
 \label{Tab:fit-nonuniversal}
 \scalebox{1.0}{
 \begin{tabular}[t]{|l|l|l|l|l|l|l|l|l|l|l|l|}
 \hline
     M.                          &   Obs.       & $p_c$             & $a$       & $b_1$            & $b_2$
  &  M.                          &   Obs.       & $p_c$             & $a$       & $b_1$            & $b_2$ \\
 \hline
 {\multirow{3}{*}{$\rm{DM}^b$}}  & $R^{(x)}$    & 0.389\,589\,22(18)  &$0.901(4)$  & ~~$0.012(13)$            &~~$0.04(17)$
 &{\multirow{3}{*}{$\rm{DM}^s$}} & $R^{(x)}$    & 0.429\,986\,96(19)  &$0.653(4)$  & ~~$0.023(9)$       &$-0.53(11)$  \\
 \cline{2-6}
 \cline{8-12}
                                 & $R^{(a)}$    & 0.389\,589\,1(1)  &$1.236(2)$  & $-0.006(6)$      &~~$0.08(7)$
 &                               & $R^{(a)}$    & 0.429\,987\,15(12)  &$0.895(2)$  & ~~$0.043(5)$     &$-0.73(6)$  \\
 \cline{2-6}
 \cline{8-12}
                                 & $R^{(3)}$    & 0.389\,589\,40(20) &$0.480(1)$  &~~$0.011(7)$       &~~$ 0.1(1)$
 &                               & $R^{(3)}$    & 0.429\,986\,81(24) &$0.3463(8)$ & $-0.001(6)$      &$-0.27(7)$ \\
 \hline
 {\multirow{3}{*}{$\rm{SC}^b$}}  & $R^{(x)}$    & 0.248\,811\,84(3) &$1.25(2)$  & ~~$ 0.001(5)$           &~~$0.29(6)$
&{\multirow{3}{*}{$\rm{SC}^s$}}  & $R^{(x)}$    & 0.311\,607\,65(5) &$0.721(5)$  & ~~$ 0.024(4)$      &$-0.44(5)$  \\
 \cline{2-6}
 \cline{8-12}
                                 & $R^{(a)}$    & 0.248\,811\,85(3) &$1.69(1)$   & $-0.011(4)$        &~~$0.78(4)$
 &                               & $R^{(a)}$    & 0.311\,607\,69(4) &$0.992(4)$  & ~~$0.036(3)$       &$0.02(3)$  \\
 \cline{2-6}
 \cline{8-12}
                                 & $R^{(3)}$    & 0.248\,811\,94(5) &$0.651(8)$  & ~~$0.004(4)$     &~~$0.03(5)$
 &                               & $R^{(3)}$    & 0.311\,607\,70(8) &$0.384(4)$  & ~~$0.002(4)$     &$-0.46(4)$  \\
 \hline
 {\multirow{3}{*}{$\rm{BCC}^b$}} & $R^{(x)}$    & 0.180\,287\,6(1) &$2.069(8)$  & $-0.006(7)$        &~~$0.2(1)$
&{\multirow{3}{*}{$\rm{BCC}^s$}} & $R^{(x)}$    & 0.245\,961\,48(6) &$1.032(7)$  & ~~$0.020(4)$       &$-0.41(5)$  \\
 \cline{2-6}
 \cline{8-12}
                                 & $R^{(a)}$    & 0.180\,287\,57(9) &$2.839(4)$  & $-0.016(4)$        &$~~0.01(5)$
 &                               & $R^{(a)}$    & 0.245\,961\,51(6) &$1.407(3)$  &~~$0.027(3)$        &$-0.46(3)$  \\
 \cline{2-6}
 \cline{8-12}
                                 & $R^{(3)}$    & 0.180\,287\,65(9) &$1.102(3)$  &  ~~$-0.004(4)$      &~~$0.27(6)$
 &                               & $R^{(3)}$    & 0.245\,961\,46(9)   &$0.543(2)$  &  ~~0.001(3)      &$-0.19(4)$  \\
 \hline
 {\multirow{3}{*}{$\rm{FCC}^b$}} & $R^{(x)}$    & 0.120\,163\,79(7) & $3.87(2)$ &~~$0.004(8)$     &~~$0.05(12)$
&{\multirow{3}{*}{$\rm{FCC}^s$}} & $R^{(x)}$    & 0.199\,235\,17(6) &$1.48(3)$  & ~~$0.011(6)$       &$-0.13(8)$  \\
 \cline{2-6}
 \cline{8-12}
                                 & $R^{(a)}$    & 0.120\,163\,80(5) & $5.311(6)$  &  $-0.008(4)$       &~~$0.04(5)$
 &                               & $R^{(a)}$    & 0.199\,235\,22(5) & $2.077(3)$  & ~~$0.018(4)$       &$-0.13(4)$  \\
 \cline{2-6}
 \cline{8-12}
                                 & $R^{(3)}$    & 0.120\,163\,72(18) & $2.059(5)$  & ~~$0.014(6)$   &$-0.1(1)$
 &                               & $R^{(3)}$    & 0.199\,235\,12(9)   & $0.804(2)$  & ~~$0.002(5)$  &~~$0.09(6)$ \\
 \hline

\end{tabular}
}
 \end{center}
 \end{table*}

\subsection{Separate fits}
In numerical study of phase transitions, dimensionless ratios like $Q_1$ and $Q_2$
are known to provide powerful tools for locating critical points $p_c$.
The wrapping probabilities have analogous finite-size scaling behaviors as the dimensionless ratios,
and thus should also provide a useful method for estimating $p_c$.
This is demonstrated in Fig.~\ref{RxQs} for site percolation on the DM lattice.
The intersections of the $Q_2$ data for different sizes $L$ would approximately give
the percolation threshold $p_c \approx 0.429 \, 95 $, with uncertainty at the fourth or fifth decimal place.
Due to their faster convergence as $L$ increases, the intersections of the $R^{(x)}$ data
would yield $p_c \approx 0.429 \, 99 $.
Similar phenomena are observed in all the percolation models studied in this work.
Thus, it clearly suggests that
the wrapping probabilities are more powerful tools for estimating $p_c$ than the dimensionless ratios $Q_1$ and $Q_2$.

According to the least-squares criterion, we fit Monte Carlo data for the
quantities $R^{(x)}$, $R^{(a)}$, $R^{(3)}$, $Q_1$ and $Q_2$
separately for each percolation model to the following scaling ansatz
 \begin{eqnarray}
U(p,L) &=&U_0+ \sum_{k=1}^3 q_k  (p-p_{c})^k L^{k y_t}  \nonumber \\
 & &+b_{1} L^{-1.2}+b_{2} L^{-2} \;,
 \label{eq:sepfit}
 \end{eqnarray}
where $y_t$ is the thermal exponent, $U_0$ is a universal value depending on the quantity studied,
and the $q_k$ ($k=1,2,3$) and $b_j$ ($j=1,2$)
are non-universal constants. A correction exponent of $-1.2$ is taken from the existing literature~\cite{WangZhouZhangTimDeng13}.
To evaluate the systematic errors caused by the scaling terms which are not included in the fitting ansatz,
we set a lower cutoff $L \ge L_{\rm min}$ on the data and study the effect on the residual $\chi^2$ as $L_{\rm min}$ increases.
Generally, we prefer the fit which produces
$\chi^2/DF \sim O(1)$ ($DF$ is the degree of freedom), and in which the
subsequent increases of $L_{\rm min}$ do not drop $\chi^2$ by vastly more than
one unit per degree of freedom. These principles apply in all
the fits we carry out.

In the fits, we try different combinations of corrections to scaling:
(1) both $b_1$ and $b_2$ are free to be determined by the data;
(2), $b_1$ is set to $0$ and $b_2$ is free; and (3), $b_1$ is free and $b_2$ is fixed at $0$.
We find that the correction amplitudes $b_1$ for the wrapping probabilities
are rather small and in many cases are statistically consistent with zero.
In contrast, for the dimensionless ratios one clearly observes a non-zero correction amplitude $b_1$.
Moreover, the amplitudes $q_1$ of the term $ q_1(p-p_c)L^{y_t}$ in Eq.~(\ref{eq:sepfit})
for $R^{(x)}$ and $R^{(a)}$ are larger than those for $Q_1$ and $Q_2$.
This suggests that the wrapping probabilities are more sensitive to the deviation from criticality $p-p_c$
than the dimensionless ratios.
These observations in the fits are reflected by Fig.~\ref{RxQs}.

Tables~\ref{Tab:sep-fit1} and~\ref{Tab:sep-fit2} summarize the percolation thresholds
and the amplitudes $q_1$ from our preferred fits with combination (1),
where the uncertainties are just the statistical errors.
It can be seen that the estimates of $p_c$ from different quantities are consistent with each other within
the combined error margins.
Further, the wrapping probabilities yield more accurate estimate of $p_c$ than the dimensionless ratios
by a factor of two or three.

\begingroup
 \begin{table}
 \begin{center}
 \caption{Simultaneous fits of the wrapping probabilities $R^{(x)}, R^{(a)}, R^{(3)}$ for all models.}
 \label{Tab:fit-universal}
 \scalebox{1.0}{
 \begin{tabular}[t]{|l|l|l|l|l|}
 \hline
   Obs.          & $y_t$       & $U_0$            & $U_2$      & $U_3$    \\
 \hline
  $R^{(x)}$      & 1.1424(11)   & $0.257\,80(6)$   &$1.23(1)$  &$-0.9(6)$   \\
  $R^{(a)}$      & 1.1418(4)   & $0.460\,02(2)$   &$0.311(2)$  &$-0.99(4)$  \\
  $R^{(3)}$      & 1.1413(6)   & $0.080\,46(4)$  &$4.98(1)$   &~~$9.7(4)$  \\
\hline
\end{tabular}
}
 \end{center}
 \end{table}
 \endgroup

\begin{figure}[htbp]
\centering
\includegraphics[scale=0.95]{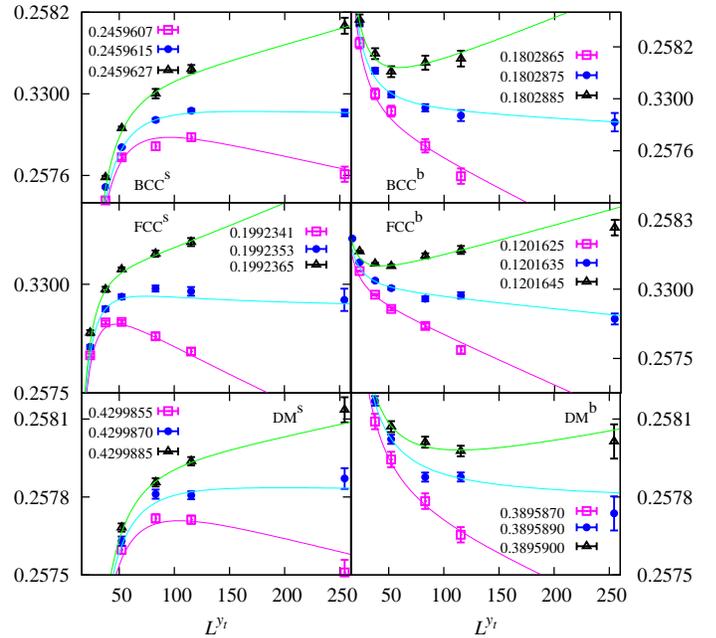}
\caption{$R^{(x)}(p,L)$ versus $L^{y_t}$ at given $p$ values which are in close to
 the estimated percolation thresholds for the site and bond percolation on the BCC (top),
 FCC (middle) and DM (bottom) respectively.}
\label{pc_judge}
\end{figure}

\begin{table*}[htbp]
\caption{Final estimates of percolation thresholds for the three-dimensional percolation models.
  The error bars include both statistical and systematic errors. }
\label{tabfinalpc1}
\begin{tabular}[t]{|l|ll|lll|}
\hline
Lattice            & Bond               &                                     &   & Site               &                   \\
                   & $p_c$(Present)     & $p_c$(Previous)                     &   & $p_c$(Present)     & $p_c$(Previous)       \\
\hline
DM            & 0.389\,589\,2(5)  & 0.389\,3(2)~\cite{Marck98}                   &   & 0.429\,987\,0(4)      &0.430\,1(4)~\cite{Marck98} \\
                   &                   & 0.390(11)~\cite{VyssotskyGordonFrischHammersley61}  &   &    &0.426(+0.08,-0.02)~\cite{SilvermanAdler90} \\
\hline
SC                 & 0.248\,811\,85(10) & 0.248\,811\,82(10)~\cite{WangZhouZhangTimDeng13}   &    & 0.311\,607\,68(15)& 0.311\,607\,7(2)~\cite{WangZhouZhangTimDeng13} \\
                   &                    & 0.248\,812\,6(5)~\cite{LorenzZiff98b}      &   &      & 0.311\,607\,4(4)~\cite{DengBlote05} \\
\hline
BCC                &0.180\,287\,62(20)     & 0.180\,287\,5(10)~\cite{LorenzZiff98a}    &     & 0.245\,961\,5(2)     & 0.245\,961\,5(10)~\cite{LorenzZiff98b}    \\
                   &    &      &          &         & 0.246\,0(3)~\cite{BradleyStrenskiDebierre91},\,0.246\,4(7)~\cite{GauntSykes83}  \\
\hline
FCC                & 0.120\,163\,77(15)   & 0.120\,163\,5(10)~\cite{LorenzZiff98a}   &    & 0.199\,235\,17(20)       & 0.199\,236\,5(10)~\cite{LorenzZiff98b} \\
\hline
\end{tabular}
\end{table*}

\subsection{Simultaneous fits}
As described above, the Monte Carlo simulations for the SC lattice are much more extensive
and are performed on larger system sizes than those on the other lattices.
This leads to the more precise estimates of $p_c$ and other parameters on the SC lattices.
It is noted that for a given wrapping probability or dimensionless ratio,
the value of $U_0$ in Eq.~(\ref{eq:sepfit}) is universal.
To make use of the extensive simulation for the SC lattice,
we carry out a simultaneous analysis of the Monte Carlo data for all the percolation systems
studied in this work.
More precisely, we choose the wrapping probabilities $R^{(x)}$, $R^{(a)}$, and $R^{(3)}$,
and for each of them, the data is fitted by
\begin{eqnarray}
 U(p_j,L) &=&U_0+ \sum_{k=1}^3 U_k a_j^k (p_j-p_{c,j})^k L^{k y_t}  \nonumber \\
 & &+b_{1,j} L^{-1.2}+b_{2,j} L^{-2} \;,
 \label{eq:simfit}
 \end{eqnarray}
where $U_k \; (k=0, 1, 2, 3)$ and $y_t$ are universal; $j \in \{ 1,2,...,8\}$
refer to the site and bond percolation models on DM, SC, BCC and FCC lattice,
and the parameters with subscript $j$ are model-dependent.
In other words, Eq.~(\ref{eq:simfit}) can be regarded as a set of
equations in which the universal parameters $U_k$ and $y_t$ are shared by all the percolation models.
We expect that an accurate estimation of these universal parameters will be mainly
achieved by the high-precision Monte Carlo data on the SC lattice,
and as in return, this will help to improve the accuracy of $p_c$ for the other models.
Such a simultaneous analysis has been applied to the 3D Ising model, and
the derivation of Eq.~(\ref{eq:simfit}) can be found in Ref.~\cite{DengBlote03}.

The simultaneous fits by Eq.~(\ref{eq:simfit}) follow the same procedure as that in the above subsection.
We first note that among $U_1$ and $a_j$ with $j =1, \cdots, 8$, there is one redundant parameter,
and we thus set $U_1=1$.

Tables~\ref{Tab:fit-nonuniversal} and~\ref{Tab:fit-universal}
summarize the results for the universal parameters, the percolation thresholds,
and other non-universal constants, taken from the preferred fits with $L_{\rm min}= 24$ or $32$.
In these fits, both the correction amplitudes $b_{1,j}$ and $b_{2,j}$ are left free.
It can be seen from Tab.~\ref{Tab:fit-nonuniversal} that
the leading correction amplitudes $b_{1,j}$ are rather small. In the cases that $b_{1,j}$
cannot be distinguished from zero within the statistical uncertainties,
one can in principle exclude the leading correction term in the fits, which will further
decrease the error margins.

In comparison with the results in Tab.~\ref{Tab:sep-fit1} from the separate fits,
the simultaneous analyses do significantly improve the estimates of $p_c$.
By taking into account the results from different wrapping probabilities and from fits with different $L_{\rm min}$,
we obtain the final estimates of $p_c$, as summarized in Tab.~\ref{tabfinalpc1}.
To check the reliability of the final quoted error margins in Tab.~\ref{tabfinalpc1},
we plot the $R^{(x)}$ data at $p_c$ and two other $p$ values which are away from $p_c$
about four or five times of final error bars.
Precisely at $p=p_c$, the $R^{(x)}$ data should tend to a horizontal line as $L \rightarrow \infty$,
whereas the data at $p \neq p_c$ will bend upward or downward.
This is indeed clearly seen in these plots, some of which are shown in Fig.~\ref{pc_judge},
confirming the reliability of our final results in Tab.~\ref{tabfinalpc1}.

Also presented in Tab.~\ref{tabfinalpc1} are existing estimates of $p_c$ from the literature.
It can be seen that this work does provide the percolation
thresholds $p_c$ with higher precision.
For the bond and site percolation models on the DM lattices, such improvement is significant.

\section{Results at $p_c$}
\label{results at pc}
By fixing $p$ at or very close to the estimated thresholds $p_c$ in Tab.~\ref{tabfinalpc1},
we study the covariances $g^{(x)}_{bR}$ and $g^{(x)}_{sR}$,
the largest-cluster size $C_1$, the shortest-path length $S$, and the cluster-number density $\rho$.
From their finite-size-scaling behaviors, one can determine the thermal and magnetic
renormalization exponent $y_t$ and $y_h$,
the shortest-path fractal dimension $\dm$, and the universal excess cluster number $b$.
In addition, we also obtain the thermodynamic cluster-number densities $\rho_c$ for the studied percolation models.

\begin{figure}
\centering
\includegraphics[scale=0.9]{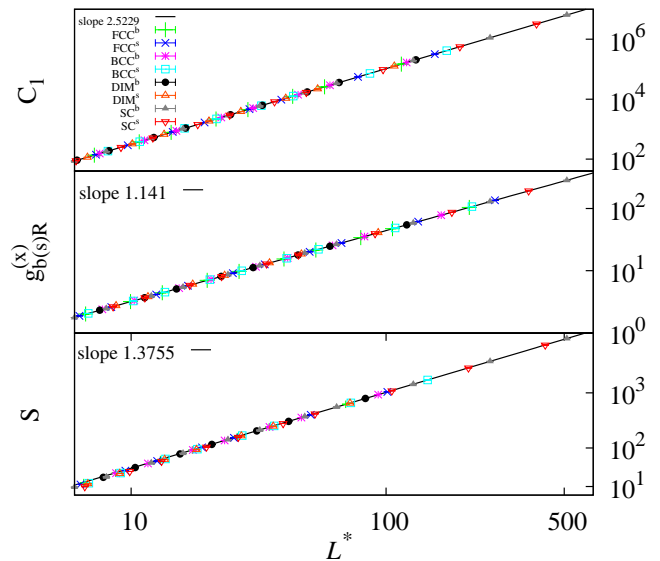}
\caption{Log-log plot of $C_1$ , $g^{(x)}_{b(s)R}$ and $S$
versus the rescaled linear size $L^* $ for all the $8$ percolation models.
We set $L=L^*$ for the bond percolation on the SC lattice, and rescale $L$ by a  constant factor (model-dependent) to collapse the numerical data.}
\label{dmin_fig}
\end{figure}

\subsection{Exponents $y_t$, $y_h$ and $\dm$}

Following an analogous simultaneous analysis procedure, we fit
the data of $g^{(x)}_{bR}$ and $g^{(x)}_{sR}$, $C_1$, and $S$ by the ansatz
 \be
 \scrA = L^{y_\scrA} ( a_{0,j} + b_{1,j}L^{-1.2} + b_{2,j}L^{-2})\; ,
 \label{eq:A}
 \ee
where ${y_\scrA}$ is the universal scaling exponent.
It is $y_t$ for covariance $g^{(x)}_{bR}$ and $g^{(x)}_{sR}$,
$y_h$ for the largest-cluster size $C_1$, and $\dm$ for the shortest-path length $S$.
We obtain $y_t = 1.141 \, 3 (15)$, $y_h = 2.522 \, 93 (10)$, and $\dm = 1.375 \, 5 (3)$,
which are consistent with the estimates in Ref.~\cite{WangZhouZhangTimDeng13}, with comparable or slightly better precision.
For an illustration of these universal exponents, we plot in the log-log scale the data of these quantities
versus the rescaled linear size $L^*=w L$, with constant $w=1$ for the bond percolation on the SC lattice.

 \subsection{Excess number of clusters}

The numerical data of the cluster-number density at percolation thershold for all the studied percolation models
are simultaneous fitted by the scaling ansatz
 \begin{eqnarray}
 \rho= \rho_c + V^{-1}(b + b_{1,j}L^{-2}) \; ,
 \label{eq:rhok}
 \end{eqnarray}
where $V$ is the number of lattice sites, and the correction amplitude $b$ is known to be also universal
and is referred to as the excess cluster number in Ref.~\cite{ZiffFinchAdamchik97}. The subleading correction is taken to be $-2$.
Due to the rapid decay of the correction term, the finite-$L$ data of $\rho$ quickly converges to
the thermodynamic value $\rho_c$; the well-determined values of $\rho_c$ then aids in estimating the correction amplitude $b$ from the small-$L$ data.
The fitting results of $\rho_c$ and $b$ are shown in Table~\ref{Tab:fit-rhok}.
Taking into account some potential systematic errors--e.g., due to the small deviation of
the simulated $p$ value from $p_c$, we have the final estimate $b = 0.675(1)$.

\begin{table}
 \caption{Simultaneous fits of $\rho$ at the thresholds. Fitting parameter $L_{\rm min}=16$ is set for
  all the models.}
 \label{Tab:fit-rhok}
 \scalebox{1.00}{
 \begin{tabular}[t]{|l|l|l|l|}
 \hline
  M.                  & $\rho_c$                   &  $b$       & $b_1$\\
 \hline
   $\rm DM^b$                          & 0.231\,953\,78(4) & {\multirow{8}{*}{$0.674\,7(4)$}}   & $-0.6(5)$  \\
 \cline{1-2}
 \cline{4-4}
   $\rm DM^s$                          & 0.075\,519\,45(2) &    & $-1.1(4)$  \\
 \cline{1-2}
 \cline{4-4}
   $\rm SC^b$                           & 0.272\,932\,836(9) &    &~~1.1(2) \\
 \cline{1-2}
 \cline{4-4}
   $\rm SC^s$                           & 0.052\,438\,217(3)&    & $-0.02(12)$  \\
 \cline{1-2}
 \cline{4-4}
   $\rm BCC^b$                          & 0.298\,343\,834(12) &    & ~~$0.3(3)$  \\
 \cline{1-2}
 \cline{4-4}
   $\rm BCC^s$                          & 0.040\,045\,144(3) &    & $-0.76(9)$  \\
 \cline{1-2}
 \cline{4-4}
   $\rm FCC^b$                          & 0.307\,691\,25(2) &    & ~~$0.1(2)$  \\
 \cline{1-2}
 \cline{4-4}
   $\rm FCC^s$                          & 0.026\,526\,453(4) &    & $-0.3(2)$  \\
 \hline
 \end{tabular}
 }
 \end{table}

\begin{figure}
\centering
\includegraphics[scale=0.46]{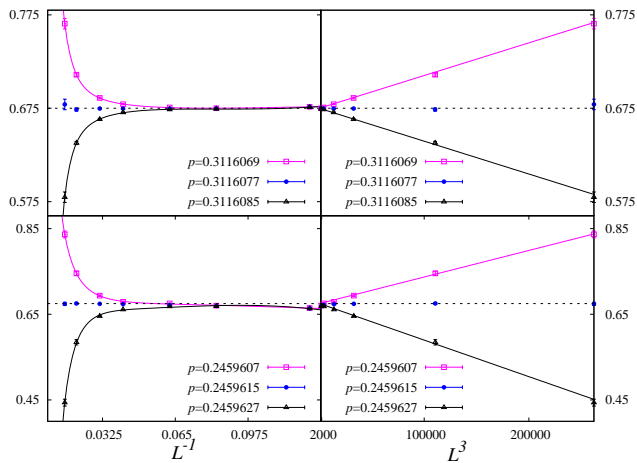}
\caption{
Excess cluster number $V(\rho-\rho_c)$ ($\equiv b$) versus $L^{-1}$(left) and $L^3$(right)
for SC site (top) and BCC site (bottom) percolation models.
The dashed straight lines represent constant $0.675$.
}
\label{rho_fig}
\end{figure}

An illustration of the excess cluster number $b$ is shown in Fig.~\ref{rho_fig},
where the values of $V (\rho-\rho_c)$ are plotted versus $1/L$ for the site percolation
on the SC and the BCC lattices.
It can be seen that the $V (\rho-\rho_c)$ values at $p=p_c$ quickly converge to $b=0.675$,
while those for $p \neq p_c$ are either bending downward or upward.
However, this does not imply that the cluster-number density $\rho$
provides a good quantity for locating $p_c$.
Near $p_c$, the finite-size behavior of $\rho(p,L)$ near threshold $p_c$
can be described by
 \begin{eqnarray}
 \rho (p, L) &=&\rho_c + f_1(p - p_c) + f_2 (p-p_c)^2 + V^{-1}[b+ \nonumber \\
    & & h_1 (p-p_c)L^{y_t}+h_2(p-p_c)^2L^{2y_t} + \cdots ] \; ,
 \label{eq:rhok_finite}
 \end{eqnarray}
where $f_i$ and $h_i$ ($i=1,2$) are non-universal parameters.
The critical density $\rho_c$ and the terms with $f_i $ 
arise from the analytical part of $\rho(p,L)$ and do not depend on size $L$.
They dominate the finite-size scaling of $\rho(p,L)$ but cannot be used to determine $p_c$.
This is illustrated in Fig.~\ref{rho_fig}.
The critical singularity is reflected in the subleading terms with $L$-dependence.
For the site percolation on the SC lattice, the fit yields
$p_c = 0.311\,604(2)$, with much larger error margin than those from wrapping probabilities.

\section{Summary}
\label{discussion}

We present a Monte Carlo study of the bond and site percolation on several three-dimensional lattices,
and obtain high-precision estimates of the percolation thresholds (Tab.~\ref{tabfinalpc1}),
the cluster density (Tab.~\ref{Tab:fit-rhok}), the wrapping probabilities (Tab.~\ref{Tab:fit-universal})
and the excess cluster number b = 0.675(1).
These accurate scientific data can serve as a testing ground
for future study of systems in the percolation universality class.
More importantly, it is observed that the wrapping probabilities can be a useful and reliable
approach for locating phase transitions.
It is very plausible that this observation generally holds
in other statistical-mechanical systems that have suitable graphical representations.

\section{Acknowledgments}
We thank R. M. Ziff and T. M. Garoni for helpful suggestions.
This research was supported in part by NSFC under Grant No. 91024026, 11275185 and 11147013, and the Chinese Academy of Science.
We also acknowledge the Specialized Research Fund for the Doctoral Program of Higher Education
under Grant No. 20103402110053.
The simulations were carried out on the NYU-ITS cluster, which is partly supported by NSF Grant No. PHY-0424082.

\end{document}